\documentclass[%
 aip, amsmath,amssymb,
reprint,]{revtex4-1}
\usepackage{graphicx}
\usepackage{dcolumn}
\usepackage{bm}

\usepackage[utf8]{inputenc}
\usepackage[T1]{fontenc}
\usepackage{mathptmx}

\usepackage[T1]{fontenc}
\usepackage[utf8]{inputenc}
\usepackage{gensymb}
\usepackage{textcomp}
\usepackage{amsmath}
\usepackage{graphicx}
\usepackage{url}
\PassOptionsToPackage{hyphens}{url}
\usepackage{hyperref}

\usepackage{color}
\definecolor{blue}{rgb}{0,0,1}

\definecolor{dblue}{rgb}{0,0,0.5}
\newcommand{\revise}[1]{\textcolor{black}{#1}}
\definecolor{dblue}{rgb}{0,0,0.5}
\newcommand{\finalRevise}[1]{\textcolor{black}{#1}}
\definecolor{dgreen}{rgb}{0,0.5,0}

\begin{document}
\title{A dynamical view on the Turkish power grid}
\title{Dynamical modelling of cascading failures in the Turkish power grid}

\author{Benjamin Sch\"afer}
\email{b.schaefer@qmul.ac.uk}
\affiliation{School of Mathematical Sciences, Queen Mary University of London, London E1 4NS, United Kingdom}
\affiliation{Chair for Network Dynamics, Center for Advancing Electronics Dresden
(cfaed) and Institute for Theoretical Physics, Technical University
of Dresden, 01062 Dresden, Germany}

\author{G. Cigdem Yalcin}
\affiliation{Department of Physics, Istanbul University, 34134, Vezneciler, Istanbul, Turkey}

\date{\today}

\begin{abstract}
A reliable supply of electricity is critical for our modern society and any large scale disturbance of the electrical system causes substantial costs. In 2015, one overloaded transmission line caused a cascading failure in the Turkish power grid, affecting about 75 million people. 
Here, we analyze the Turkish power grid and its dynamical and statistical properties. 
\revise{Specifically, we propose, for the first time, a model that incorporates the dynamical properties and the complex network topology of the Turkish power grid to investigate cascading failures.}
We find that the network damage depends on the load and generation distribution in the network with centralized generation being more susceptible to failures than a decentralized one. Furthermore, economic considerations on transmission line capacity are shown to conflict with stability.
\end{abstract}

\pacs{}

\maketitle 

\begin{quotation}
Cascading failures in power grids are the main cause of large-scale blackouts, which cause large economic and social costs. Previous studies mainly modeled cascades as a sequence of steady states, employing mostly static analysis. However, real-world cascading failures often commence within short time scales such as few minutes \revise{or} even seconds, rendering static analysis inappropriate. Here, we focus on the Turkish power grid, which experienced a major blackout in 2015. We extract the Turkish power grid topology from Transmission System Operator (TSO) data  and introduce a framework to describe dynamical cascading failures.
\end{quotation}

\section{Introduction}
We are surrounded by natural or man-made networks, either technologically or sociologically that directly or indirectly affect human beings and are studied extensively as complex systems \cite{Boccaletti2006,Latora2017}.
Many advancements of our modern society rely on a stable electricity supply and thereby on the network of electrical power grids. Examples range from medical care to long-distance transportation, instant communication or industrial automation, most of which break down without electricity \cite{bundestag2011ta}. Therefore, we can clearly identify the electrical supply system  as \emph{uniquely critical} \cite{Obama2013}. Any large scale outage will cause high economic costs and potentially threaten the political stability of a region \cite{Bialek2007}.

The electricity system is based on a high-voltage alternating current (AC) network, the power grid. This electrical grid cannot store any energy in itself \cite{Kund94}, in contrast to,  for example,  the gas network \cite{Osiadacz1987}. Instead, supply and demand have to be balanced at all times. Any imbalance will cause a shift of the power grid's frequency away from the reference of $50$~Hz (or $60$~Hz in the Americas and parts of South-East Asia). A shortage of generation reduces the frequency and an abundance of generation increases the frequency, speeding up the turbines at the connected power plants \cite{Mach08}.
Any large deviation from the reference frequency will cause disconnection of loads (load shedding) and shutdown of power plants, either to keep the grid stable or to protect the machines \cite{Mach08}. Simultaneously, transmission lines are protected by automatic shut-down mechanisms, which disconnect the line if the load on the line exceeds a given security threshold, i.e., the line trips.

Cascading failures are sequences of events where the disruption of a single element causes a global breakdown (blackout) of the system \cite{Crucitti2004,Crucitti2004a,Kinney2005,Ji2016}.
Suppose a single element of the power grid fails, e.g. a large generator has to shut down, a major load is disconnected or an important transmission line is lost. The removal of this one element then induces a disturbance in most of the remaining network elements. In the worst case, this disruption triggers a  domino-like cascade of errors where additional lines become overloaded and trip and also generators have to disconnect \cite{Dobson2007}.

Despite the overall improvements in supply security, we are still facing multiple major blackouts worldwide every decade. Constraining ourselves to major power blackouts that affected more than twenty million people at least, we obtain the following list: Venezuelan 2019 (30 millions), Sri Lanka 2016 (21 millions), Kenya 2016 (44 millions), Turkey 2015 (75 millions), Pakistan 2015 (140 millions),  Bangladesh 2014 (150 millions), India 2012 (620 millions), Indonesia 2005 (100 millions), United States-Canada 2003 (55 millions), Italy-Switzerland 2003, (55 millions), Philippines 2002, (40 millions), India 2001 (230 millions), Philippines 2001, (35 millions) \cite{wikipedia}.
One of the more recent extreme events in this list is the major power blackout of Turkey in 2015 that affected the entire population of the country, not only parts of the country, like some of the other examples.

Turkey's \finalRevise{geography} is quite unique with major cities in its Western European region, while a large area of its land is also on the Asian continent. Due to this prominent geographical feature, the relationship between power subsystems of the Western and Eastern parts of Turkey also corresponds to an obvious intercontinental link between Europe and Asia as well. 
Turkey's power grid is connected to the Continental European power grid, operated by the ENTSO-E (European Network of Transmission System Operators) and experienced a major blackout in 2015. Contrary to cascading events in Europe in 2003 and 2006 \cite{Bialek2007,UCTE_Report2007}, this contingency was limited to Turkey and did not spread to the remaining ENTSO-E grid as the transmission lines between the Turkish and the remaining European grid were disconnected. 
Interestingly, very few studies have investigated the blackout and the cascading failure in Turkey in 2015 \cite{Veloza2016,Shao2016}. Hence, we briefly summarize the sequence of events, which left about 75 million people without electrical power \cite{Hurriyet2015,Hurriyet2015a}:

On 31st March 2015, preceding its largest power outage in recent history, Turkey's main generation and main demand regions were spatially separated. Due to large amounts of water flowing through the hydro power plants in the Eastern parts of Turkey, most power plants in the West were generating very little electricity. Simultaneously, the large cities in the West, such as Izmir and Istanbul had a large demand for electricity to operate industrial and household applications. Therefore, the lines connecting the East and the West were heavily loaded. During the morning, one line connecting both regions tripped as it became overloaded. 
Consequentially, the flow from East to West was re-routed on other East-West connections, which then also became overloaded.  Thereby,  the Western grid completely disconnected from the Eastern one and lost about 4.7 GW of power, missing approximately $20\%$ of the necessary power to fulfill the current demand \cite{Turkish2015}. This shortage of power resulted in a rapid drop of the Western grid's frequency. Loads were disconnected to stabilize the grid. Still, several generators started to trip and therefore continued to destabilize the grid. After about 10 seconds, the Western system fully collapsed  \cite{Turkish2015}. Simultaneously, the Eastern grid had a considerable surplus of electricity generation, resulting in a similar loss of synchrony and thereby a simultaneous collapse within seconds.

A dynamical description of the power grid is necessary to capture the brief time scale of seconds that passed from the initial failure until a full collapse of the grid.
Fortunately, the interest in modelling power grids dynamically has risen substantially in recent years \cite{12powergrid,Lozano2012,13powerlong,14bifurcation,Menck2014,Schmietendorf2014,Nish15,Schaefer2018}.
Similarly, cascading failures have been studied intensively in statistical physics and engineering \cite{Simonsen2008,Hines2010,Buldyrev2010,Brum13,Koc2014,Pahwa2014,Witthaut2015,
Plietzsch2016,Rohden2016,Manik2017,Ronellenfitsch2017,Cetinay2017,guo2019complex}. Still, combining the two approaches, i.e, using a dynamical framework for cascading failure analysis is still rare but received additional support very recently \cite{Simonsen2008,song2015dynamic,Yang2017b,schaefer2018Cascade,huang2019small}.

Here, we first extract an approximate representation of the Turkish power grid topology. We then introduce our dynamical framework for the  short time scale dynamics of cascading failures, which is based on the swing equation. Next, we analyse the Turkish grid topology using this framework to find decentralized generator distributions slightly more favourable than centralized ones. Finally, we point out how economic optimization and grid resilience compete when considering network extensions.

\begin{figure*}
\begin{centering}
\includegraphics[width=1.7\columnwidth]{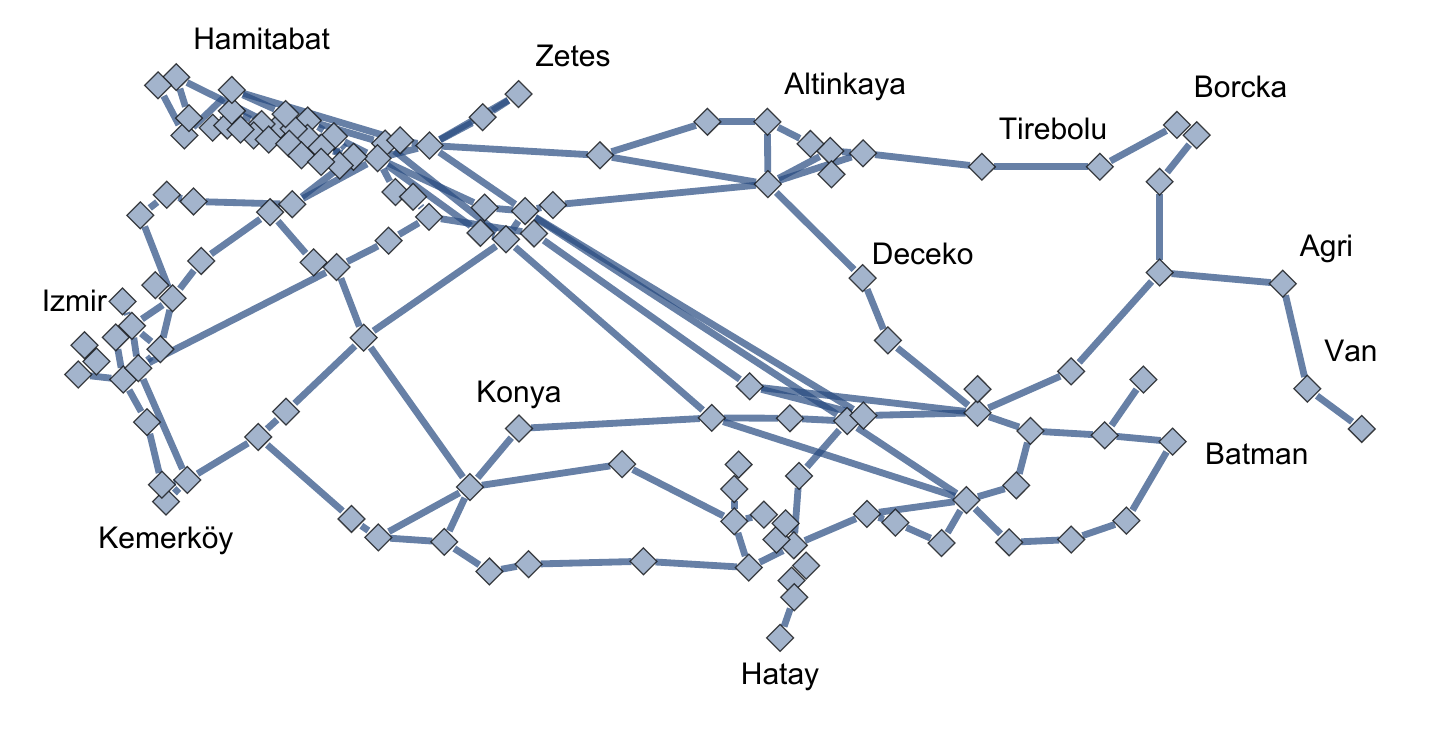}
\par\end{centering}
\caption{\textbf{Approximate topology of the Turkish power grid, based on ENTSO-E data \cite{Turkish2015}.} The grid has many connections
and substations close to its major cities in the West (Istanbul, Ankara, Izmir etc). In contrast, the more sparsely populated Eastern part is dominated by hydro power plants and long-distance transmission lines.
\label{fig:TurkishGrid}}
\end{figure*}

\section{The Turkish grid}
\label{sec:TrukishGrid}

We require a network representation of the Turkish power grid to investigate and analyze any (dynamical) phenomenon, such as the blackout that occurred in 2015. Therefore, we extract the approximate topology using publicly available data from the report on the blackout by ENTSO-E  \cite{Turkish2015} and the interactive ENTSO-E map \cite{ENTSOE_MAP}. 
The maps mark the position of substations, which we treat as the nodes of the Turkish network, and transmission lines, which form the edges of the Turkish grid. Details of the line parameters or the precise demand and generation at each node are not available to us. For our dynamical model below, we will assume different random realizations to model the various states of the grid. Considering multiple scenarios grants deeper insight into the power grid's behavior as demand and generation distribution will typically be different in summer than in winter, similarly as night demand profiles differ from ones during the day \cite{Yang2017a}.

Inspecting the extracted Turkish power grid, we notice a large cluster of substations close to the highly populated regions, e.g. close to Istanbul in the North-West of Turkey, see Fig. \ref{fig:TurkishGrid}. In contrast, the central and Eastern parts of Turkey are less populated and also have fewer substations.
These subjective observations are \revise{backed-up} by network measures \cite{Newman2010}: With $N=127$ nodes and $|E|$=174 edges the Turkish grid topology is a sparse network with a highly connected and highly clustered community in the North-West and a very sparse and almost tree-like structure to the East.

The extracted topology, including the names of the substations and their approximate positions, is freely available for download, see \revise{Appendix: Methods}.
With the grid topology available, we formulate the dynamical model determining the state of each node (substation) in the next section.

\section{A dynamical power grid model}

\begin{figure}
\begin{centering}
\includegraphics[width=0.9\columnwidth]{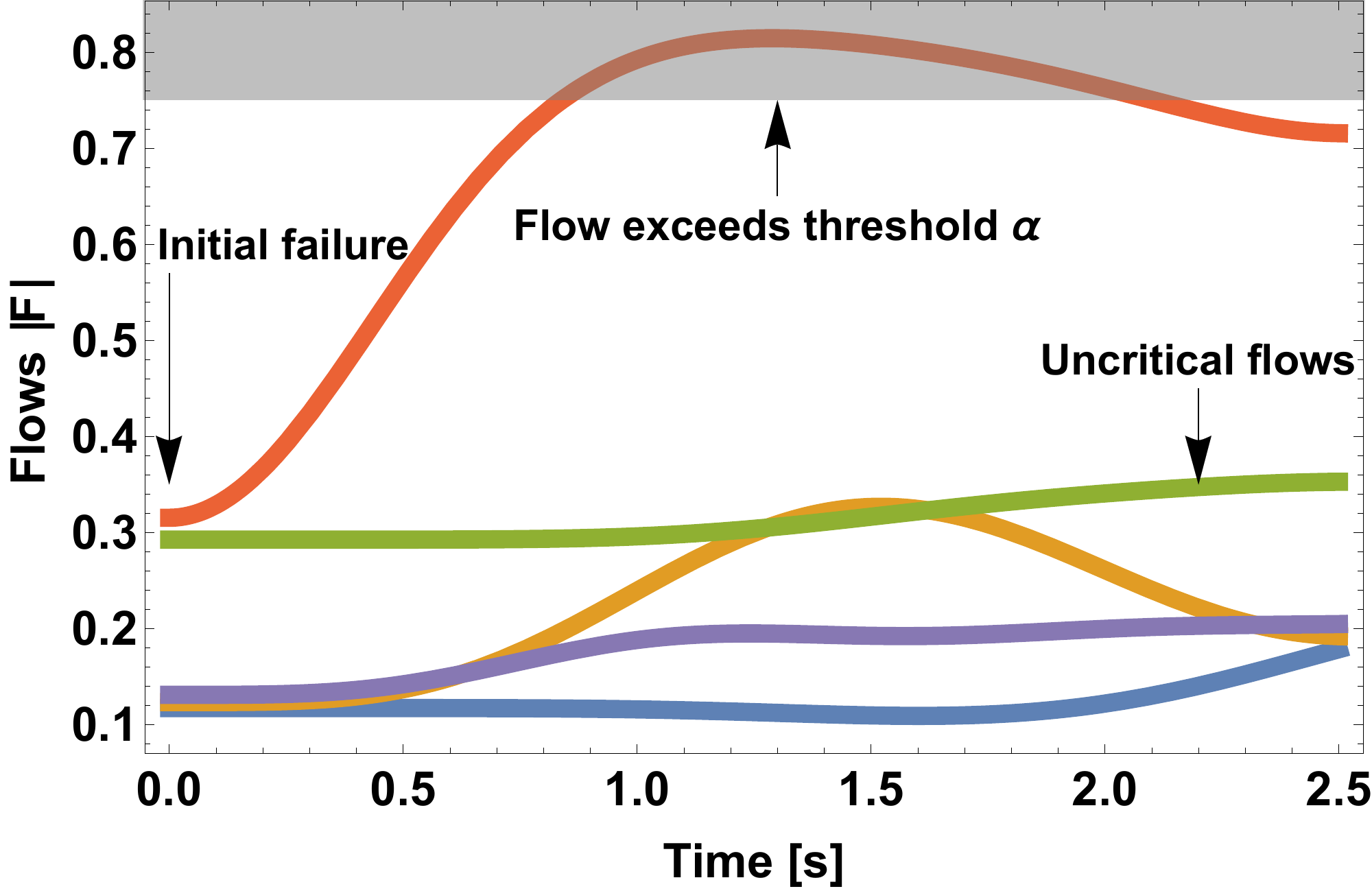}
\par\end{centering}
\caption{\textbf{\revise{Following an initial failure, the flow of few critical lines exceeds a threshold.}} We display the flows along 5 lines representative for the flows in the network. The initial conditions of the simulation use the  fixed point  of the original full network. Then, we let the most heavily loaded line trip, resulting in a dynamical transition towards a new fixed point. This transition involves large transient power flows on several lines, which might violate security thresholds $\alpha $ dynamically, while most line flows remain uncritical. \label{fig:Cascade_Flow_Trajectory}}
\end{figure}

To model line failures within the time scale of seconds, we formulate a dynamical model capturing the essential properties of the power grid dynamics. Each node in our approximation of the Turkish grid is a substation in the high voltage transmission grid. Modelling all underlying voltage structures, machines and devices would result in a very complex model, which would not allow easy structural and analytical insights. Therefore, we will focus on the transmission level of the power grid and make a couple of simplifying assumptions: Since the voltage on the transmission lines is typically between 220 and 380 kV, we neglect any ohmic losses and reactive power, thereby we only consider active power transmission. The voltage amplitude is assumed to be constant as well. Furthermore, we  aggregate all consumers and generators connected to each substation into one effective synchronous machine. This machine acts as an effective generator on the transmission grid if it is connected to many generators, e.g. hydro or fossil fuel plants. Contrary, nodes in urban areas will often act as effective consumers.

Mathematically, we make use of the well-known swing equation \cite{Fila08,Mach08,Nish15}.
Each node $i\in\left\{ 1,...,N\right\} $ is then described by its deviation from the reference or mains frequency $f_R=50$ Hz or $60$ Hz. We express the deviation in terms of the angular velocity $\omega_i=2\pi \left(f_i- f_R\right)$ in the rotating frame, i.e., $\omega_i=0$ is equivalent to a frequency of $f_i=50$~Hz (or $60$ Hz).

The other state variable of each node is the voltage phase angle $\theta_i$, which crucially determines the power flow between two nodes. Overall the swing equation can be expressed as follows
\begin{eqnarray}
\frac{\text{d}}{\text{d}t}\theta_{i} & = & \omega_{i},\label{eq:Swing equation}\\
\frac{\text{d}}{\text{d}t}\omega_{i} & = & P_{i}-\gamma\omega_{i}+\sum_{j=1}^{N}K_{ij}\sin\left(\theta_{j}-\theta_{i}\right),
\end{eqnarray}
with effective power $P_i$ (positive for generators and negative for consumers), damping constant $\gamma_i$ and coupling \begin{equation}
K_{ij}=B_{ij}V_{i}V_{j},
\end{equation} where $B_{ij}$ is the susceptance between two nodes and $V_i$ is the voltage amplitude \cite{Kund94}. 
We often use homogeneous coupling $K_{ij}=K  A_{ij}$ with an unweighted adjacency matrix $A_{ij}$ and then only need to fix the coupling constant $K$.  \revise{Note that high voltage transmission lines use three-phase currents, while we are using a simplified single-phase calculation that assumes all three phases to be symmetrical.}

In its steady state or fixed point, all angles $\theta_i$ are constant and all angular velocities $\omega_i$ are zero. Contrary, if the system is undergoing a perturbation, e.g. the loss of a transmission line, the dynamical nature of the system becomes clear.

Let us consider the Turkish grid to rest at its stable state and suddenly a line is lost, e.g. due to an overload, a lightening strike or similar. Then, the flows $F_{ij}$
on many lines will change. Especially, if the lost line was carrying a large current, changes will be substantial. While some flows stay almost constant or decrease, other lines will have to carry the flow of the lost line. Eventually, the system might settle down at a new fixed point but in the transition period from one steady state to the next one, overload criteria on the transmission lines might be exceeded, see Fig. \ref{fig:Cascade_Flow_Trajectory}.

We link the dynamical swing equation with considerations of cascading failures as follows. Throughout all simulations, we track the \revise{relative} flow $F(t)$, given by
\begin{equation}
F_{ij}(t)=\sin(\theta_j(t)-\theta_i(t))
\end{equation}  as a function of time. 
Most transmission lines will be disconnected automatically if they exceed a security threshold $\alpha$ \cite{Crucitti2004,Crucitti2004a}. We will be strict in our security measures and assume a line $(i,j)$ is disconnected as soon as $F_{ij}(t)>\alpha$. \finalRevise{Note that $F_{ij}$ gives the relative fraction of maximal power flow along a line, which would be reached if $\sin\left(\theta_j(t)-\theta_i(t)\right)=1$. The threshold $\alpha=0.5$ means that a line is assumed to overload if it is transmitting 50\% of this maximal load, regardless of its absolute physical capacity.}



Our simulations work as follows: Without knowledge of the precise distribution of generation and consumption throughout the network, we assume randomly distributed generators and consumers so that the total generation matches the total consumption.
We consider three different scenarios using the same grid topology as introduced in Section \ref{sec:TrukishGrid}:

First, we consider many small distributed generators and consumers with a total of 63 consumers with consumption $P^\text{Con}=-1s^{-2}$ and 64 generators with generation $P^\text{Gen}=\frac{63}{64}s^{-2}\approx 1s^{-2}$.
We use homogeneous coupling of $K=4.5s^{-2}$ on all lines.

Secondly, we model a more centralized generation pattern with fewer generators, namely 18 generators with $P^\text{Gen}=\frac{109}{18}s^{-2}\approx 6s^{-2}$. The remaining 109 nodes are consumers, still with $P^\text{Con}=-1s^{-2}$. This centralized power distribution requires a larger coupling $K$ to reach its stable state \cite{Rohden2016}. Hence, we use  a homogeneous coupling of $K=12s^{-2}$.

Finally, we emulate a more economical investment into the grid infrastructure. We initiate the grid at the same parameters as in the first case, i.e., homogeneous coupling and many small generators. Then, we change the coupling $K_{ij}$ on all lines so that it is \revise{approximately twice as large as the relative flow $F_{ij}$. Since a change in the network coupling $K_{ij}$ changes the flows $F_{ij}$, we iterate this procedure a couple of times until most lines carry about $50\%$ of their maximum capacity.}  We call this case heterogeneous coupling since all lines might have a different coupling constant $K_{ij}$. See also \cite{schaefer2018Cascade} for details and \revise{Appendix:Methods} for a link to the applied coupling matrix. \revise{Note that these heterogeneties are on the network topology level, i.e. different edges now have different properties. These are still synthetic transmission line values and each individual line in itself is considered physically homogeneous and we also still assume that the three-phase voltage can be simplified as a single-phase voltage.}

Before we started the simulations, we prepared each network in its fixed point. At $t=1$ second, we remove one trigger line $(a,b)$ from the network. Then, we update the flows $F_{ij}(t)$ over time and cut other lines as soon as their flow exceeds the threshold $F_{ij}>\alpha$. We consider all lines in the network as potential trigger lines. The threshold $\alpha$ is chosen so that at the fixed point and time $t=0$ seconds no line is overloaded, i.e. $F_{ij}(0)<\alpha$ on all lines $(i,j)$.
Cascade computations are done for up to  50 seconds of simulation time, during which all observed cascades terminated.

\section{Cascading failures in the Turkish power grid}

\begin{figure*}
\begin{centering}
\includegraphics[width=1.9\columnwidth]{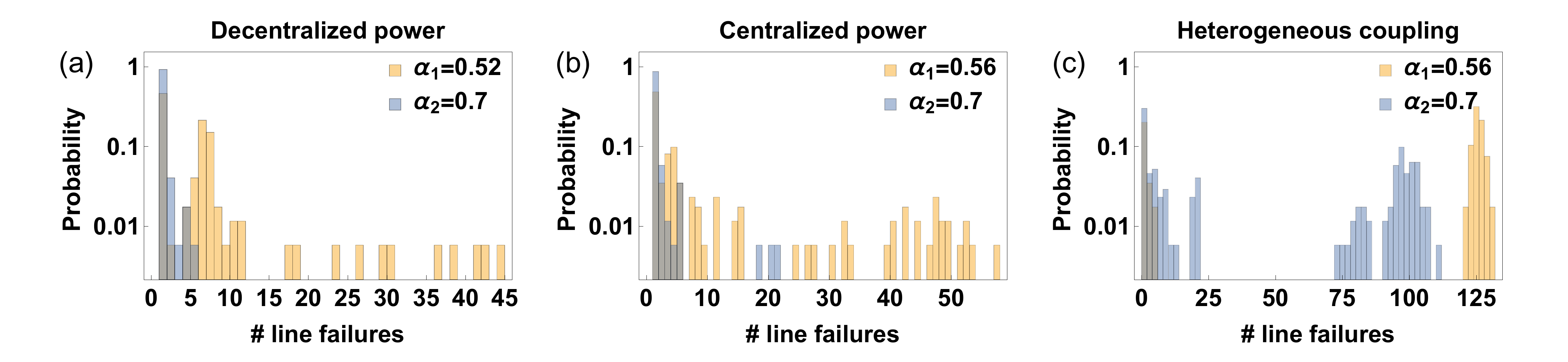}
\par\end{centering}
\caption{\textbf{Heterogeneous coupling leads to many outages.} \revise{We consider each line of the graph as a trigger line and then record the number of lines failures. While some lines lead to no additional failures, others lead to many line failures. This histograms give the probability for a certain number of lines to fail  when randomly choosing an initial trigger line in the different network configurations.} The total number of edges and thereby maximum damage is $|E|$=174.   (a): Many small distributed generators, with homogeneous coupling $K=4.5s^{-2}$, lead to mostly small cascades. (b): Fewer, but larger generators, also with homogeneous coupling $K=12s^{-2}$, show similar statistics. (c): Same power distribution as in (a) but using heterogeneous coupling so that $F_{ij}\approx0.5$ on all lines. This results in the largest cascading failures. \label{fig:CascadeStatistics}}
\end{figure*}

\begin{figure*}
\begin{centering}
\includegraphics[width=1.99\columnwidth]{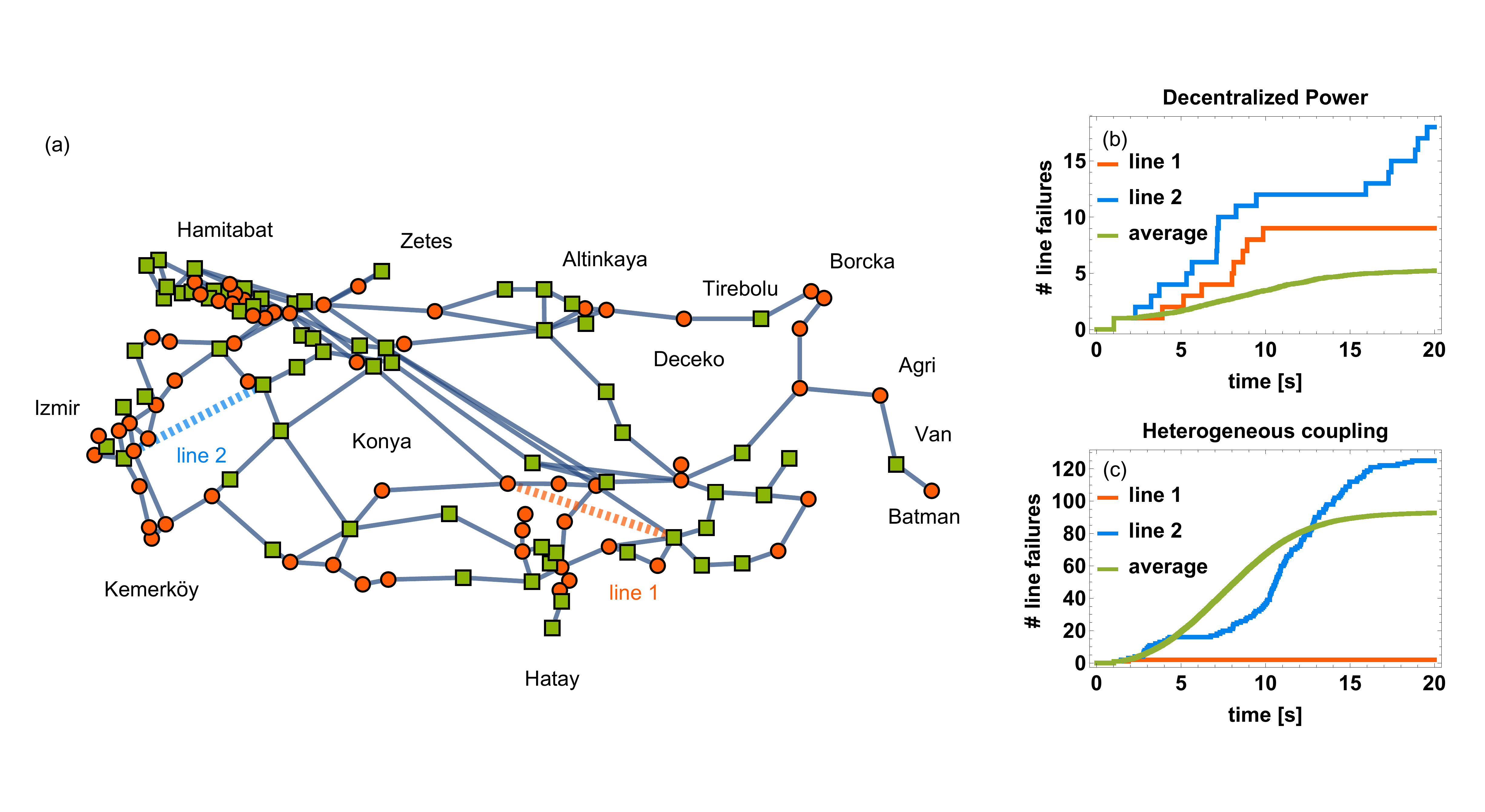}
\par\end{centering}
\caption{\textbf{The expected network damage highly depends on the position of the initial failure.} (a): We consider the Turkish power grid topology with randomly placed generators (green squares) and consumers (red circles) with power $P\approx \pm1s^{-2}$. Following two scenarios, we either cut line 1 (orange) or line 2 (blue). (b): Depending on which line failed, a different number of total failures is recorded in the  homogeneously coupled grid. (c): The position of the initial failure is particularly important for the statistics of the heterogeneously coupled grid. While line 1 results in almost no failures, an initial failure of line 2 results in a complete collapse of the grid. Furthermore, the number of failures of an individual event and the average number of failures (green) is much higher than in the homogeneous case. Note that we only display failures up to 20 seconds, at which point most cascades terminated. To not miss any later events, we simulate until 50 seconds. \label{fig:TriggerLineDependenFailures}}
\end{figure*}

\begin{figure*}
\begin{centering}
\includegraphics[width=1.75\columnwidth]{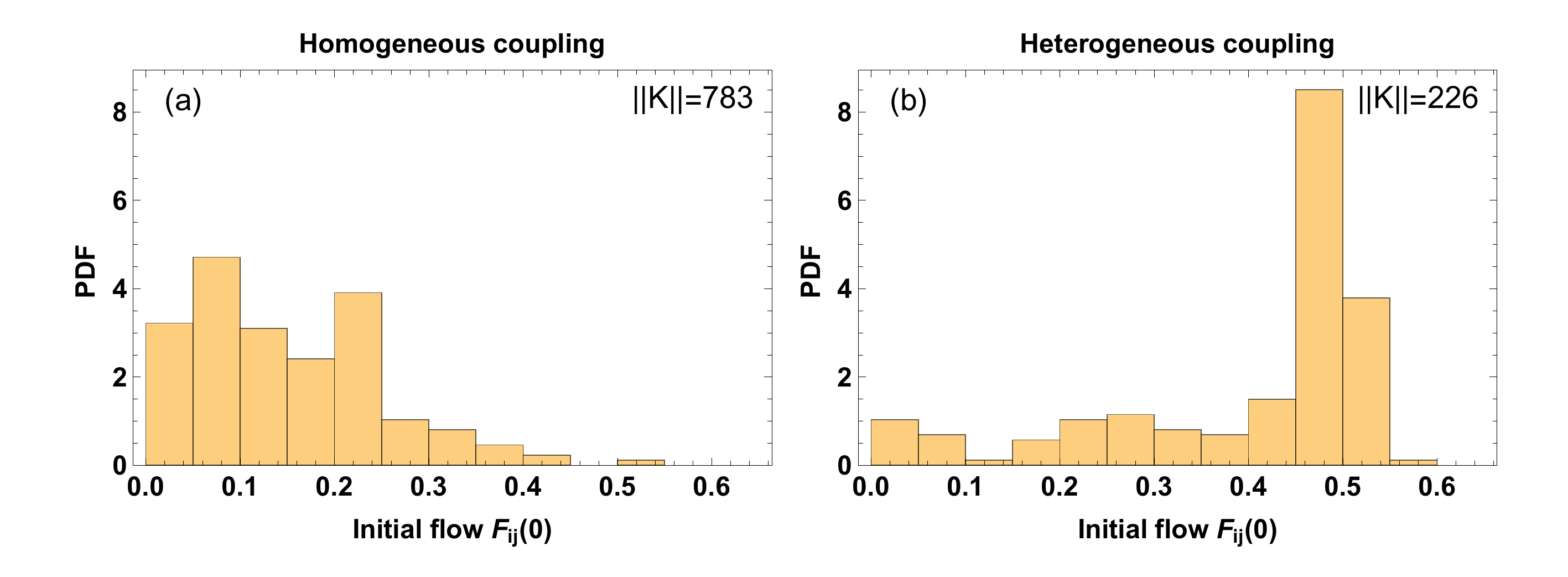}
\par\end{centering}
\caption{\textbf{The heterogeneously coupled grid has more highly-loaded lines.} We display the probability density function (PDF) for all lines in the network to have a certain initial load $F_{ij}(0)=|\sin(\theta_j(0)-\theta_i(0))|$, where the angles $\theta(0)$ are fixed point solutions of the swing equation \eqref{eq:Swing equation}. The homogeneously coupled grid (a) has many lines with very small loads, while the heterogeneously coupled grid (b) has many highly-loaded lines. \label{fig:InitialFlowDistribution}}
\end{figure*}

We now apply the dynamical cascading framework to investigate the statistical properties of cascading failures in the Turkish grid topology. We start by asking: How many lines will fail if one random line were to trip? We answer this question in a structured way by considering all \revise{lines} in the network as a potential trigger line $(a,b)$. Then, we record how many lines did fail at the end of each simulation and compare two different tolerances $\alpha$. Aggregating the results for all trigger lines yields a histogram of the number of line failures, see Fig.~\ref{fig:CascadeStatistics}.

For a given network, a higher tolerance $\alpha$ reduces the expected damage on the network substantially. While for $\alpha_1=0.52$ the decentralized power topology can have cascading events with up to 45 failed lines, a higher tolerance of $\alpha_2=0.7$ restricts the maximum network damage to approximately 5 lines (Fig.~\ref{fig:CascadeStatistics} a).
We observe a very similar behavior of decreasing network damage with increasing tolerance for centralized power and heterogeneous coupling (Fig.~\ref{fig:CascadeStatistics} b \& c).

Interestingly, we do note that especially for homogeneous coupling many lines do not cause any cascade (Fig.~\ref{fig:CascadeStatistics} a \& b), and the most likely case is to observe 1 line failure, i.e., only the trigger line $(a,b)$ failed. Contrary, heterogeneous coupling results in many large cascades, even for the higher tolerance of $\alpha_2=0.7$ many lines cause cascades of 75 failures or more (Fig.~\ref{fig:CascadeStatistics} c).

To understand the different cascade responses in more detail, we investigate the number of line failures as a function of time in Fig. \ref{fig:TriggerLineDependenFailures}. We compare two different trigger lines as highlighted in Fig. \ref{fig:TriggerLineDependenFailures} a and record the aggregated number of line failures over time for the decentralized, homogeneous coupling (Fig. \ref{fig:TriggerLineDependenFailures} b) and the heterogeneous coupling (Fig. \ref{fig:TriggerLineDependenFailures} c). Again, we notice that the heterogeneous coupling results in many more line failures than the decentralized setting.
After a brief period of a few seconds with no or few failures, the system experiences a large number of contingencies until either most lines already tripped or the grid eventually stabilizes so that no further lines are lost.
While \emph{line 1} does not lead to any secondary line failures beyond the trigger line failure, \emph{line 2} causes large damage in the heterogeneous case, see Fig. \ref{fig:TriggerLineDependenFailures} c. The cascading events in Turkey in 2015 were likely set in motion by a failure of such a "\emph{line 2}"\revise{, i.e., a line critical for the operation of the stable state either because it is carrying a large load or because alternative rerouting pathways are already highly loaded \cite{16critical}.} This also emphasizes that some lines are critical for the network's operation while the failure of other lines can be more easily compensated.

Why do heterogeneously coupled grids display such large cascades? To answer this, we compare the flow distribution of the homogeneously and the heterogeneously coupled grids at their respective fixed points in Fig. \ref{fig:InitialFlowDistribution}. 
By construction, the heterogeneous coupling has many lines with a \revise{relative} load of $F\approx 0.5$ and lines with very low \revise{absolute} load are likely not essential to operate the grid. Contrary, the homogeneously coupled grid has many lines with low load because the coupling on all lines has to be increased so that the highest loaded line is still within security margins. Thereby, the homogeneously coupled grid is more robust with respect to line failures as its lines carry a lower average flow of $\left<F^\text{Homo. coupling}\right>\approx 0.15$, compared to  $\left<F^\text{Hetero. coupling}\right>\approx 0.4$ of the heterogeneously coupled grid.

However, the high robustness of the homogeneously coupled grid is expensive. The total coupling necessary, i.e. $||K||=\frac{1}{2}\sum_{i=1}^N \sum_{j=1}^N K_{ij}$, is about 3 times larger in the homogeneous case than in the heterogeneous case. \revise{This  intuitively makes sense as the total power transmitted $P^\text{Trans.}$ within the network should be about the same between the two cases. This transmitted power is given approximately as the product of the coupling and the relative flow $P^\text{Trans.}\approx ||K||\left<F\right> $.}
We conclude that robustness has to be paid by investments into the grid infrastructure.

\section{Discussion}

Motivated by the fast time scale of the Turkish blackout in 2015, we have introduced a dynamical description of cascading failures. 
We have extracted an approximation of the Turkish power grid's topology based on data from the transmission system operators \cite{Turkish2015,ENTSOE_MAP}. Furthermore, we discussed the swing equation as a means to model the short time scale of the power grid dynamics and presented a framework for cascading failures \cite{schaefer2018Cascade}.

Crucially, both the network topology and the simulation framework are available for download and further usage, see the link in the \revise{Appendix: Methods}.
Thereby, we offer a tool for other scientists to investigate additional questions related to the Turkish grid or any instance of cascading failures in any power grid. 
The Turkish power grid is of special interest as it bridges Europe and Asia and a failure within the Turkish grid disconnects the Continental interconnection.
Also, the technique presented here \cite{schaefer2018Cascade} may be used to understand other major power blackouts by investigating the impact of network structure and dynamical characteristics of the power grids on the blackout dynamics. 

For the Turkish case, we found that whether a specific line failure will cause no, a small or a large-scale cascade depends critically on the distribution of generation and demand as well as on the line properties throughout the network. We highlighted this by comparing centralized and decentralized generation schema as well as considering heterogeneous coupling.
Interestingly, the distribution of generation and demand within a given grid is not static but changes over time, e.g. due to seasonal changes, night and day differences etc. \cite{Yang2017a}. Even cultural routines and geographical properties of the regions may effect the distribution of generation and demand of the power grid, e.g. based on cooking habits, air condition usage etc. Consequently, the resilience of the grid has to be monitored continuously, as it is done in modern power grids \cite{Mach08}. 
Our contribution of the dynamical framework \cite{schaefer2018Cascade} could be a starting point for dynamical properties to be integrated into the monitoring.
Furthermore, weighting resilience with respect to cascading failures has to be balanced with economic considerations, as we have shown in Fig. \ref{fig:InitialFlowDistribution}. Investing in few strategically important lines may lead to larger cascades than a more spread out investment in transmission lines. Still, the former is considerably cheaper and more likely to be accepted by local communities \cite{ciupuliga2013role}. \revise{These economic and social considerations further add to previous insights that not every added transmission lines is necessarily beneficial for the grid's stability \cite{witthaut2012braess,tchuisseu2018curing}.}

Our dynamical framework is based on the swing equation \cite{Kund94,Fila08} with several simplifications (e.g. constant voltage amplitude, no ohmic losses or reactive power transmission). We believe these simplifications are justified in order to have access to the dynamical properties and to allow (semi-)analytical investigations, e.g. to predict critical lines \cite{schaefer2018Cascade} or observe propagation patterns. Tools for more detailed analysis of power systems already exist \cite{ElectricTestCases, Birchfield2017} but only allow specific case studies instead of a systematic investigation. 

Many aspects of cascading failures are still not fully explored. The most pressing question would be how to mitigate cascading failures. If we were to anticipate the fast transient events, could we disconnect specific regions of the grid to enable a steady operation of the remaining grid? Can we identify topologies and generation patterns that are more robust than others? In addition, we could expand our cascading framework to explicitly allow generators connected to substations to disconnect, thereby changing the node properties dynamically without altering the network topology.
\revise{Finally, real power grids are very heterogeneous systems and these heterogeneous properties heavily influence stability properties \cite{wolff2018power} and could even reduce overload risks \cite{schiel2017resilience}. An interesting and ideal future project would be to test our predictions by using real node and edge properties.}

\begin{acknowledgments}
B. Sch{\"a}fer gratefully acknowledges support from the Federal Ministry of Education
and Research (BMBF grant no. 03SF0472A-F and 03EK3055F) and
the German Science Foundation (DFG) by a grant toward the Cluster
of Excellence ``Center for Advancing Electronics Dresden'' (cfaed). This project has received funding from the European Union’s Horizon 2020 research and innovation programme under the Marie Sklodowska-Curie grant agreement No 840825. 
G. C. Yalcin was supported by the Scientific Research Projects Coordination Unit of Istanbul University with project number 32990. G. C. Yalcin also gratefully acknowledges to the Asian Network of Complexity Scientists (ANCS), Complexity Institute, Nanyang Technological University.

\includegraphics[width=0.4\textwidth]{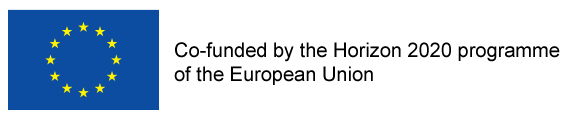}
\end{acknowledgments}

\appendix*

\section{Methods}
The details of the simulations, including all network parameters, topology of the Turkish grid and simulation examples are available at \url{https://osf.io/gd5xn/}.

\bibliography{TurkishGrid}

\end{document}